\documentclass[a4paper,aps,reprint,pra, 11pt, onecolumn, superscriptaddress,nofootinbib, showpacs, notitlepage]{revtex4-1} 
\usepackage{times}
\usepackage{array}
\usepackage{amsmath}
\usepackage{amsfonts}
\usepackage{amssymb}
\usepackage{amsthm}
\usepackage{amsbsy}
\usepackage{stmaryrd}
\usepackage{pifont}
\usepackage{latexsym}
\usepackage{mathtools}
\usepackage{bbold}
\usepackage{gensymb}
\usepackage[all]{xy}
\usepackage{graphicx}

\usepackage{epstopdf}
\usepackage{epsfig}
\usepackage[labelformat=default, labelsep=default, justification=centerlast,  font=footnotesize]{caption}
\usepackage{verbatim}
\usepackage{multirow}
\usepackage{rotating}
\usepackage{minibox}

\usepackage[titletoc,toc,title]{appendix}
\usepackage{enumerate}
\usepackage{enumitem}
\usepackage{color}
\usepackage[dvipsnames]{xcolor}
\usepackage[right=2.75cm, left=2.75cm, top=3cm, bottom=3cm]{geometry}

\usepackage{comment}
\usepackage[colorlinks]{hyperref}
\usepackage{sidecap}
\usepackage{bbm}
\usepackage{bm}

\DeclareCaptionJustification{justified}{\justifying}
\begin{document}

\title{Relativistic quantum reference frames: the operational meaning of spin}
\author{Flaminia Giacomini}
\author{Esteban Castro-Ruiz}
\author{\v{C}aslav Brukner}
\affiliation{Vienna Center for Quantum Science and Technology (VCQ), Faculty of Physics, University of Vienna, Boltzmanngasse 5, A-1090 Vienna, Austria}
\affiliation{Institute of Quantum Optics and Quantum Information (IQOQI),Austrian Academy of Sciences, Boltzmanngasse 3, A-1090 Vienna, Austria}
\date{\today}

\begin{abstract}
	
	The spin is the prime example of a qubit. Encoding and decoding information in the spin qubit is operationally well defined through the Stern-Gerlach set-up in the non-relativistic (i.e., low velocity) limit. However, an operational definition of the spin in the relativistic regime is missing. The origin of this difficulty lies in the fact that, on the one hand, the spin gets entangled with the momentum in Lorentz-boosted reference frames, and on the other hand, for a particle moving in a superposition of velocities, it is impossible to ``jump'' to its rest frame, where spin is unambiguously defined. Here, we find a quantum reference frame transformation corresponding to a ``superposition of Lorentz boosts,'' allowing us to transform to the rest frame of a particle that is in a superposition of relativistic momenta with respect to the laboratory frame. This enables us to first move to the particle's rest frame, define the spin measurements there (via the Stern-Gerlach experimental procedure), and then move back to the laboratory frame. In this way, we find a set of ``relativistic Stern-Gerlach measurements'' in the laboratory frame, and a set of observables satisfying the spin $\mathfrak{su}(2)$ algebra. This operational procedure offers a concrete way of testing the relativistic features of the spin, and opens up the possibility of devising quantum information protocols for spin in the special-relativistic regime.
\end{abstract}

\maketitle

\section{Introduction}

The description of physical systems is standardly given in terms of coordinates as defined by reference frames. Thanks to the principle of covariance, stating the equivalence of all descriptions regardless of the choice of the reference frame, it is possible to choose the reference frame where the relevant dynamical quantities can be most conveniently described. For example, it is typically easier to describe the dynamics of a system from the point of view of its rest frame, because only internal degrees of freedom contribute to the dynamics in the rest frame. 

When the external degrees of freedom (momentum) of the system are in a quantum superposition from the perspective of the laboratory, no classical reference frame transformation can map the description of physics from the laboratory to the rest frame. However, this can be achieved via a quantum reference frame (QRF) transformation between two frames moving in a superposition of velocites relative to one another. In order to achieve such change of quantum reference frame in the nonrelativistic regime, a formalism was introduced in Ref.~\cite{QRF} to change the description to a refence frame which is in a quantum relationship with the initial one. This QRF transformation only depends on relational quantities, and it has also been derived starting from a gravity inspired symmetry principle in a perspective neutral model \cite{perspective1, perspective2} . An immediate consequence of the formalism is that entanglement and superposition are QRF-dependent features. This formalism naturally leads to the possibility of identifying the rest frame of a quantum system in an operational way. 

Here, we further develop this approach in the case of a relativistic quantum particle with spin, with the goal of finding an operational description of the spin in a special-relativistic setting. Spin is operationally defined in the rest frame of a particle (or, to a good approximation, for slow velocities) via the Stern-Gerlach experiment. When the particle has relativistic velocities, the spin degree of freedom transforms in a momentum-dependent way. If a standard Stern-Gerlach measurement is performed on a particle in a pure quantum state moving in a superposition of relativistic velocities, the operational identification of the spin fails, because no orientation of the Stern-Gerlach apparatus returns an outcome with unit probability. This happens because, as shown in Ref.~\cite{peres_first}, the reduced density matrix of the spin degree of freedom is mixed when a Lorentz boost is performed and the momentum is traced out. The question arises whether it is possible to find `covariant measurements' of the spin and possibly momentum, which predict invariant probabilities in different Lorentzian reference frames also for the case of a quantum relativistic particle moving in a superposition of velocities. In this case, it would be possible to map the unambiguous description of spin in the rest frame of the particle to the frame of the laboratory, and therefore derive the corresponding observables to be measured in the laboratory frame to verify spin with probability one.

The question of finding such covariant measurements is motivated by the ubiquitous applications where the spin degree of freedom is used as a qubit, to encode and transmit quantum information. Such protocols are no longer valid in a relativistic context, thus limiting the range of applicability of techniques involving spin as a quantum information carrier. It is then important to explore possible alternative methods which could overcome this limitation. In the context of relativistic quantum information, this question has been extensively discussed \cite{peres_first, peresterno_review, alsingmilburn, gingrichadami, leeyoung_entanglementboost, bartlettterno_encoding, vedral_paradox, esteban, friis_relativisticentanglement, caban_reducedspin, vedral_operational, palmer_sterngerlach, markusphilipp_lorentz, terno_localisation} in relation to Wigner rotations \cite{wigner1939, weinberg, sexlurbantke} and has been related to the problem of identifying a covariant spin operator. The problem of identifying such covariant spin operator has arisen long before the birth of relativistic quantum information, and dates back to the early times of quantum mechanics \cite{thomas_precession, frenkel_spin, bargmannOnFrenkel}. Since then, a multitude of relativistic spin operators have been proposed \cite{bauke_relativisticspin}, such as the Frenkel \cite{frenkel_spin}, the Pauli-Luba\'{n}ski \cite{lubanski, ryder_paulilubanski, terno_tworoles}, the Pryce \cite{pryce_spin}, the Foldy-Wouthuysen \cite{foldywouthuysen, caban_spinobservable}, the Czachor \cite{czachor_spin} the Fleming \cite{fleming_spin} the Chakrabarti \cite{Chakrabarti_spin}, and the Fradkin-Good \cite{fradkingood_spin} spin operators. A comparative description of spin observables can be found in Ref.~\cite{terno_localisation}.

Here, we introduce `superposition of Lorentz boosts' which allow us to ``jump'' into the rest frame of a relativistic quantum particle even if the particle is {\it not} in a momentum eigenstate. In the rest frame, the spin observables fulfill the spin $\mathfrak{su}(2)$ algebra (the algebra of a qubit) and are operationally defined through the Stern-Gerlach experiment. We transform the set of spin observables in the rest frame back to an isomorphic set of observables in the laboratory frame. The transformed observables are in general entangled in the spin and momentum degrees of freedom. The set fulfills the $\mathfrak{su}(2)$ algebra and is operationally defined through a `relativistic Stern-Gerlach experiment': we construct the interaction and the measurement between the spin-momentum degrees of freedom and the electromagnetic field in the laboratory frame which gives the same probabilities as the Stern-Gerlach experiment in the rest frame. This set of observables in the laboratory frame allows us to partition the total Hilbert space into two (highly degenerate) subspaces corresponding to the two outcomes ``spin up'' and ``spin down''. Hence, with QRFs techniques the relativistic spin can effectively be described as a qubit in an operationally well-defined way.

\section{A relativistic Stern-Gerlach experiment}

\begin{figure}
	\begin{center}
		\includegraphics[scale=0.3]{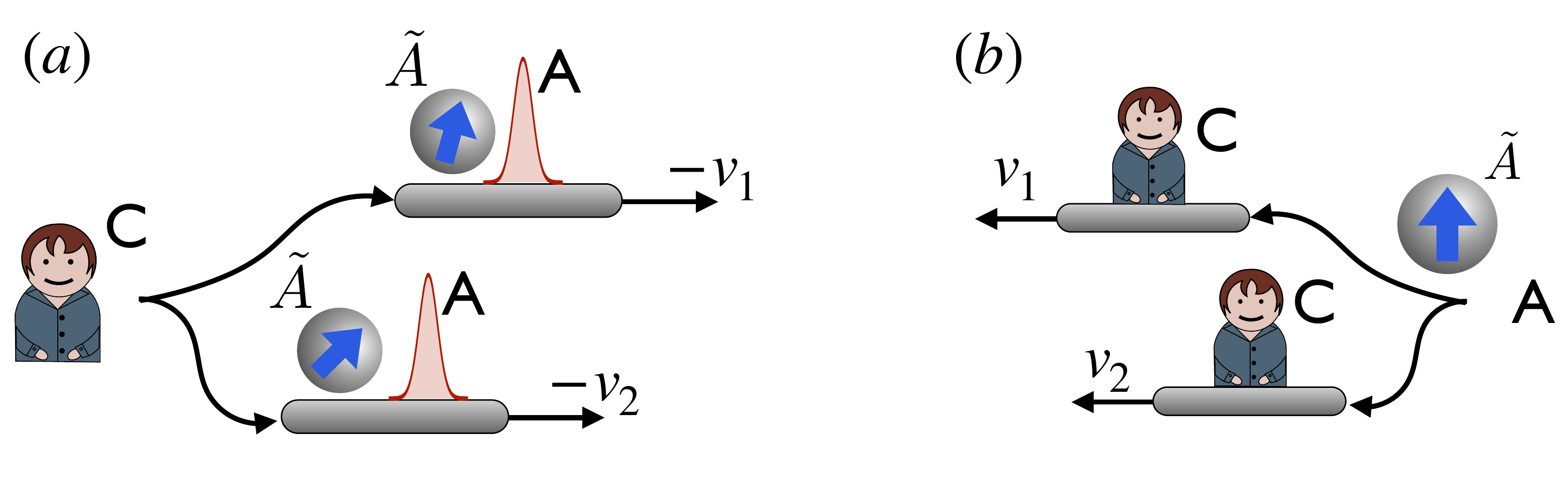}
		\caption{(a) The state of a Dirac particle A with spin $\tilde{A}$ as seen from the laboratory perspective (C). When the state is in a superposition of relativistic velocities $-v_1$ and $-v_2$, the spin degree of freedom and the momentum degree of freedom are no longer separable. (b) The state of the spin $\tilde{A}$ and of the laboratory C as seen in the rest frame of the quantum particle A. In this quantum reference frame, the spin is operationally defined by means of the Stern-Gerlach experiment.}
		\label{fig:spinQRF}
	\end{center}
\end{figure}

In the following, we build a QRF transformation between the reference frame of a laboratory C of mass $m_C$ and the rest frame of the external degrees of freedom A of a relativistic quantum particle of mass $m_A>0$ with spin degrees of freedom $\tilde{A}$, as illustrated in Fig.~\ref{fig:spinQRF}. We allow the particle to have any quantum state, and in particular to move in a superposition of momenta. This implies that there is a non-classical relationship between the initial and the final reference frame, i.e., that the rest frame A and the laboratory frame C are not related by a standard boost transformation. We show in this section how to generalise the boost transformation to this case. Formally, the situation we consider can be described by taking the one-particle sector of the positive-energy solutions of the Dirac equation\footnote{For simplicity, we only consider spin-$1/2$ particles, but the method can be straightfowardly applied to arbitrary spin.} in the Foldy-Wouthuysen representation \cite{foldywouthuysen}. 

Following Ref.~\cite{QRF} (see Supplemental Information for a review of the original formalism), when we ``stand'' in the rest frame of a particle, we describe all the systems external to the particle, but not the external degrees of freedom (i.e., the momentum degrees of freedom) of the particle itself. Hence, the quantum state describes the relational information in a given reference frame. In the reference frame in which A is at rest, the quantum state is assigned to the internal degrees of freedom $\tilde{A}$ and the laboratory  C. For simplicity, we consider that the particle and the laboratory are moving with constant, yet not necessarily well-defined, relative velocity and define the $x$ axis along the direction of the relative motion. The total state of the spin and the laboratory is assumed to be
\begin{equation} \label{eq:stateA}
	\left| \Psi \right>_{\tilde{A}C}^{(A)} = \left|\vec{\sigma} \right>_{\tilde{A}}\left| \psi \right>_{C},
\end{equation}
where $\left|\vec{\sigma} \right>_{\tilde{A}}$ is any vector representing the state of the spin in the rest frame A. In the rest frame, the spin state can in principle be tomographically verified by performing a series of standard Stern-Gerlach measurements. The state of the laboratory has a momentum-basis representation along the $x$ direction (at this stage, we neglect the quantum state in the $y$ and $z$ direction) $\left| \psi \right>_{C} = \int d\mu_C(\pi_C) \psi (\pi_C) \left| \pi_C \right>_{C}$, where $d\mu_C(\pi_C) =  \frac{d\pi_C}{(2\pi)^{1/2}\sqrt{2(m_C^2 c^2 + \pi_C^2)}}$ is the Lorentz-covariant integration measure. 

We now construct the transformation corresponding to the ``superposition of Lorentz boosts'' to the QRF of the laboratory. The unitary operator to boost to the QRF C is
\begin{equation} \label{eq:QRFTransf}
	\hat{S}_L = \mathcal{P}_{CA}^{(v)} U_{\tilde{A}}(\hat{\pi}_C),
\end{equation}
 where $U_{\tilde{A}}(\hat{\pi}_C)$ is a unitary transformation acting on the total Hilbert space $\mathcal{H}_{\tilde{A}}\otimes \mathcal{H}_C$ (notice that $\hat{\pi}_C$ is an operator), and $\mathcal{P}_{AC}^{(v)}$ is the `generalised parity operator' introduced in Ref.~ \cite{QRF}, whose explicit expression is $\mathcal{P}_{CA}^{(v)}= P_{AC} \exp \left(\frac{i}{\hbar}\log \sqrt{\frac{m_A}{m_C}}(\hat{q}_C \hat{\pi}_C + \hat{\pi}_C \hat{q}_C)\right)$, where $P_{AC}$ is the parity-swap operator mapping $\hat{x}_A \rightarrow -\hat{q}_C$ and $\hat{p}_A \rightarrow -\hat{\pi}_C$ (and viceversa), where $\hat{q}_C, \,\hat{\pi}_C$ are canonically-conjugated one-particle operators of C in the reference frame of A and $\hat{x}_A, \,\hat{p}_A$ are canonically-conjugated one-particle operators of A in the reference frame of C. Additionally to the action of $P_{AC}$, the operator $\mathcal{P}_{AC}^{(v)}$ rescales the momentum of A by the ratio of the masses of A and C, i.e., $\mathcal{P}_{AC}^{(v)} \hat{p}_A \mathcal{P}_{AC}^{(v)\dagger} = - \frac{m_A}{m_C} \hat{\pi}_C$. This enforces the physical condition that the velocity of A is mapped to the opposite of the velocity of C via the transformation\footnote{For a relativistic particle the relation between the $i$-th velocity component and the momentum is $v_i = \frac{p_i}{m_i} \left( 1+ \frac{|\vec{p}|^2}{m_i^2 c^2} \right)^{-1/2}$, where $|\vec{p}|^2$ is the norm of the spatial momentum. Therefore, only the ratio between momentum and mass determines the velocity.}. The operator $\hat{S}_L$ can be defined via its action on a basis of the total Hilbert space of the spin and the laboratory $\hat{S}_L |\vec{\sigma} \rangle_{\tilde{A}}|\pi\rangle_C = | -\frac{m_A}{m_C}\pi; \Sigma_{\pi}\rangle_{A\tilde{A}}$, where the state $|p; \Sigma_{p}\rangle_{A\tilde{A}}$ is defined via a standard Lorentz boost $\hat{U}(L_p)$ from the rest frame as $|p; \Sigma_{p}\rangle_{A\tilde{A}}=\hat{U}(L_p)| k; \vec{\sigma}\rangle_{A\tilde{A}}$ and $k= (mc, \vec{0})$ is the momentum in the rest frame. 
  In Supplemental Information we derive the transformation $\hat{S}_L$ in terms of standard Lorentz boosts connecting two relativistic reference frames where the parameter of the boost transformation is promoted to an operator. 
 
 The state of $A$ and $\tilde{A}$ expressed in the laboratory frame is $\left| \Psi \right>_{A\tilde{A}}^{(C)} = \hat{S}_L \left| \Psi \right>_{\tilde{A}C}^{(A)}$, and is explicitly written as
\begin{equation} \label{eq:StateTrans}
	\left| \Psi \right>_{A\tilde{A}}^{(C)} = \int d\mu_A(p_A) \psi\left(-\frac{m_C}{m_A}p_A\right) \left| p_A; \Sigma_{p_A} \right>_{A\tilde{A}},
\end{equation}
where $d\mu_A(p_A) =  \frac{d p_A}{(2\pi)^{1/2}\sqrt{2(m_A^2 c^2 + p_A^2)}}$ and the spin degree of freedom cannot be separated anymore from the momentum degree of freedom, which means that the state is not a product state in the laboratory frame. Notice that the effect of the $\hat{S}_L$ transformation is to apply the usual boost transformation conditional on C's momentum degree of freedom. In the laboratory frame C, unless particle A is in a sharp momentum state, no spin measurement in a standard Stern-Gerlach experiment would give a result with probability one, because of two reasons: the spin and momentum are no longer separable, and the relation between the laboratory and the rest frame is not a standard (classical) reference frame transformation. Our goal is to devise a different measurement in the laboratory reference frame, possibly involving both the spin and momentum degrees of freedom, which gives the same probability distribution as a standard Stern-Gerlach would give, if performed in the rest frame.

In order to devise such measurement we note that, in the laboratory frame, it is possible to define the observables corresponding to the spin operators in the rest frame by transforming the spin, as defined in the rest frame, with a QRF transformation
\begin{equation}
	\hat{\Xi}_i = \hat{S}_L ( \hat{\sigma}_i \otimes \mathbb{1}_C) \hat{S}_L^\dagger, \qquad i=x,y,z.
\end{equation}
In terms of the momenta and of the manifestly covariant Pauli-Luba\'{n}ski operator $\hat{\Sigma}_{\hat{p}_A} = (\hat{\Sigma}^0_{\hat{p}_A}, \vec{\hat{\Sigma}}_{\hat{p}_A})$, the operators $\hat{\Xi}_i$ are expressed as (see Supplemental Information) $\vec{\hat{\Xi}} = \vec{\hat{\Sigma}}_{\hat{p}_A} - \frac{\hat{\gamma}_A}{\hat{\gamma}_A +1}\left(\vec{\hat{\Sigma}}_{\hat{p}_A} \cdot \vec{\hat{\beta}}_A  \right) \vec{\hat{\beta}}_A$, where $\hat{\gamma}_A = \sqrt{1 + \frac{\hat{p}_A^2}{m_A^2 c^2}}$ and $\vec{\hat{\beta}}_A = \left( \hat{\beta}_A^x, \hat{\beta}_A^y, \hat{\beta}_A^z \right)$, where each component is $\hat{\beta}_A^i = \frac{\hat{p}_A^i}{\sqrt{m_A^2 c^2 + \vec{\hat{p}}_A^2}}$ with $i= x, y, z$. The operators $\hat{\Xi}_i$ are equivalent to the Foldy-Wouthuysen \cite{foldywouthuysen} or Pryce spin operator \cite{pryce_spin}. By definition, these operators satisfy the $\mathfrak{su}(2)$ algebra $\left[ \hat{\Xi}_i, \hat{\Xi}_j \right] = i \epsilon_{ijk} \hat{\Xi}_k$, and have the same eigenvalues as the Pauli operators $\hat{\sigma}_i$, $i=x,y,z$. This last property can be easily checked by choosing an eigenvector $| \lambda_i \rangle$ of the operator $\hat{\sigma}_i$ in the rest frame A, such that $\hat{\sigma}_i | \lambda_i \rangle = \lambda_i | \lambda_i \rangle$ and by noting that $\hat{\Xi}_i \hat{S}_L | \lambda_i \rangle_{\tilde{A}}| \psi \rangle_{C} = \lambda_i \hat{S}_L | \lambda_i \rangle_{\tilde{A}}| \psi \rangle_{C}$. Hence, it is possible to partition the total Hilbert space $\mathcal{H}_A \otimes \mathcal{H}_{\tilde{A}}$ into two equivalence classes, defined as
\begin{subequations}
	\begin{equation}
		\mathcal{H}_0 = \left\lbrace | \Psi \rangle_{A\tilde{A}} \in \mathcal{H}_A \otimes \mathcal{H}_{\tilde{A}}\,\, \text{s.t.}\,\, | \Psi \rangle_{A\tilde{A}} \sim \hat{S}_L | 0 \rangle_{\tilde{A}}| \psi \rangle_{C}, \forall \, | \psi \rangle_{C} \in \mathcal{H}_C \right\rbrace,
	\end{equation}
	\begin{equation}
		\mathcal{H}_1 = \left\lbrace | \Phi \rangle_{A\tilde{A}} \in \mathcal{H}_A \otimes \mathcal{H}_{\tilde{A}}\,\, \text{s.t.}\,\, | \Phi \rangle_{A\tilde{A}} \sim \hat{S}_L | 1 \rangle_{\tilde{A}}| \phi \rangle_{C}, \forall \, | \phi \rangle_{C} \in \mathcal{H}_C \right\rbrace,
	\end{equation}
\end{subequations}
where $| 0 \rangle_{\tilde{A}}$ and $| 1 \rangle_{\tilde{A}}$ are the eigenvectors of $\hat{\sigma}_z$\footnote{Notice that we could have chosen any other Pauli operator to define this partition.} and two states are said to be equivalent, i.e., $|\Psi \rangle_{A\tilde{A}} \sim \hat{S}_L | i \rangle_{\tilde{A}}| \psi \rangle_{C}$, with $i=0,1$, if they are both eigenvectors of the $\hat{\Xi}_z$ operator with the same eigenvalue. We can then build a partition of the Hilbert space into two highly degenerate subspaces, one corresponding to the ``spin up'' and the other to the ``spin down'' eigenvalue, and on which it is possible to define a set of operators satisfying the $\mathfrak{su}(2)$ algebra, which can be used to encode or decode information of a single qubit.

\begin{figure}
	\begin{center}
		\includegraphics[scale=0.3]{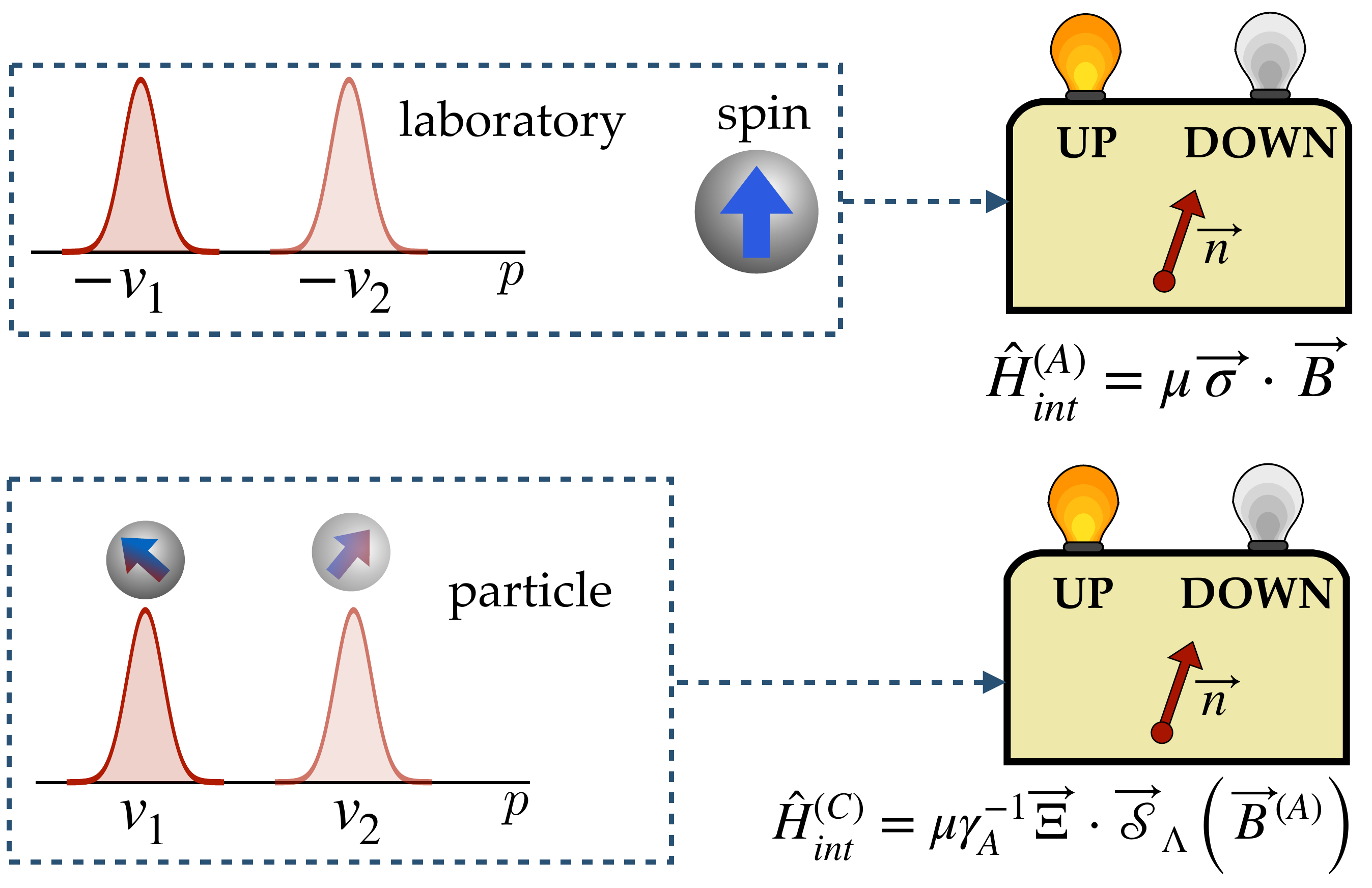}
		\caption{The relativistic Stern-Gerlach experiment as seen from the QRF A (above) and from the QRF C (below). In the rest frame of particle A, the spin is operationally defined via the Stern-Gerlach experiment. To measure spin along direction $\vec{n}$ the spin (Pauli operator) $\vec{\sigma}$ is coupled to an inhomogeneous magnetic field oriented along $\vec{n}$. The particle is then deflected towards the direction $\vec{n}$ and $-\vec{n}$ corresponding to outcome ``spin up'' and ``spin down'' respectively. When transforming to the laboratory frame C, the magnetic field and the spin transform with a superposition of Lorentz boosts for $v_1$ and $v_2$. The interaction Hamiltonian is also transformed, giving rise to a coupling between the transformed vector $\vec{\mathcal{S}}_\Lambda(\vec{B}^{(A)}) = \hat{\gamma}_A \left[ \vec{B}^{(C)} - \frac{\hat{\gamma}_A}{\hat{\gamma}_A+1}\left(\vec{\hat{\beta}}_A \cdot \vec{B}^{(C)}\right)\vec{\hat{\beta}}_A + \left(\vec{\hat{\beta}}_A \times \vec{E}^{(C)}\right) \right]$ aligned in the same direction $\vec{n}$ as the magnetic field in the rest frame, and the transformed spin operator $\vec{\Xi}$. The particle is again deflected either to $\vec{n}$ or $-\vec{n}$ corresponding to the outcome ``spin up'' and ``spin down'' respectively. The probability of detecting the outcomes ``spin up'' and ``spin down'' is preserved under change of QRF.}
		\label{fig:SternGerlach}
	\end{center}
\end{figure}

The operators $\vec{\hat{\Xi}}$ in general act on both the external and the internal degrees of freedom of the particle. Operationally, they can be defined via a ``relativistic Stern-Gerlach experiment,'' illustrated in Fig.~\ref{fig:SternGerlach}. Traditionally, in a Stern-Gerlach experiment, the spin measurement is performed by applying a magnetic field, which interacts with the spin as $\vec{B} \cdot \vec{\sigma}$ and is inhomogeneous along the direction of its orientation, i.e.,  $\vec{B} = B(\vec{r}\cdot \vec{n})\vec{n}$, where $\vec{n}$ gives the direction and $\vec{r} = (x, y, z)$. If the magnetic field is aligned precisely in the direction in which the spin state is prepared, the outcome is obtained with certainty. However, if the particle carrying the spin is moving in a superposition of relativistic velocities, no measurement of the spin alone in the laboratory frame will return the result with probability one in general. To treat such a case we set up a hypothetical Stern-Gerlach experiment in the rest frame of the particle, where the interaction Hamiltonian is $H_{int}^{(A)} = \mu \vec{B}^{(A)} \cdot \vec{\sigma}$ and $\mu$ is a coupling constant. We assume that the direction in which the magnetic field is aligned $\vec{n}$ is orthogonal to the direction of the boost $x$. Formally, this geometric configuration requires to enlarge the Hilbert space of the laboratory to the $z$ direction, which we identify with the direction $\vec{n}$ of deflection, and modify our previous definition of the state in Eq.~\eqref{eq:stateA} as $|\psi \rangle_C = |\psi_x \rangle_C |\psi_z \rangle_C$, where $|\psi_x \rangle_C$ transforms with $\hat{S}_L$ and $|\psi_z \rangle_C$ is left invariant by the transformation $\hat{S}_L$, except for the fact that the label is changed from C to A, i.e., $\hat{S}_L|\psi_z \rangle_C = |\psi_z \rangle_A$. Additionally, we assume that the motion in the $z$ direction is non relativistic. We then transform the Hamiltonian to the laboratory frame via the QRF transformation $\hat{S}_L$. Knowing that the magnetic field transforms under superposition of Lorentz boosts as $\vec{\hat{\mathcal{S}}}_\Lambda(\vec{B}^{(A)}) = \hat{\gamma}_A \left[ \vec{B}^{(C)} - \frac{\hat{\gamma}_A}{\hat{\gamma}_A+1}\left(\vec{\hat{\beta}}_A \cdot \vec{B}^{(C)}\right)\vec{\hat{\beta}}_A + \left(\vec{\hat{\beta}}_A \times \vec{E}^{(C)}\right) \right]$, we find that the interaction Hamiltonian $H_{int}^{(A)}$ is transformed to
\begin{equation} \label{eq:SternGerlachHamiltonian}
	H_{int}^{(C)} = \mu \hat{\gamma}_A^{-1} \vec{\hat{\mathcal{S}}}_\Lambda(\vec{B}^{(A)}) \cdot \vec{\hat{\Xi}}.
\end{equation}

It is straightforward to check that the direction of $\vec{\hat{\mathcal{S}}}_\Lambda (\vec{B}^{(A)})$ is also $\vec{n}$, therefore the deflection of the particle in the laboratory frame happens in the same direction as in the rest frame. Notice that, since both the quantum state and the observables transform unitarily, probabilities are automatically conserved after the change of QRF. In particular, if in the rest frame of the particle A the Stern Gerlach measurement detects that the spin is ``up'' with probability one, the ``relativistic Stern-Gerlach'' experiment in the laboratory frame with the interaction Hamiltonian of Eq.~\eqref{eq:SternGerlachHamiltonian} will also detect ``spin up'' with probability one. Note that the specific form of the electromagnetic field in Eq.~\eqref{eq:SternGerlachHamiltonian} is not crucial to our result, but we can design the coupling between the particle and the electromagnetic field according to our experimental capabilities in each reference frame. However, it is crucial that the electromagnetic field couples to the operator $\vec{\hat{\Xi}}$, unlike in the standard Stern-Gerlach experiment. In Supplemental Information, we set up a different experiment, where we couple an ihnomogenous magnetic field in the laboratory frame to give an explicit analysis of a relativistic Stern-Gerlach experiment.

It is worth noting that the interaction Hamiltonian of Eq.~\eqref{eq:SternGerlachHamiltonian} is covariant, because the quantity $H^{0} : =\hat{\gamma}_A H_{int}^{(C)}$ transforms like the zero-component of a $4$-vector. Therefore, the Schr\"{o}dinger equation in the reference frame of A, $i \hbar \frac{d}{dt_A} \left| \psi \right>_{\tilde{A}C}^{(A)} = H_{int}^{(A)}\left| \psi \right>_{\tilde{A}C}^{(A)}$, where $t_A$ is the proper time in the rest frame of A, is mapped to $i \hbar \frac{d}{dt_C} \left| \psi \right>_{\tilde{A}A}^{(C)} = H_{int}^{(C)}\left| \psi \right>_{\tilde{A}A}^{(C)}$, where $t_C$ is the proper time in the rest frame of C and the relation $t_C = \hat{\gamma}_A t_A$ holds. The general, manifestly covariant expression of  $H^0$ is
\begin{equation}
	H^{0} = \frac{1}{2}\eta^{0 \rho}\epsilon_{\rho \mu \nu \lambda} \hat{\Sigma}_{p_A}^{\mu}F^{\nu \lambda},
\end{equation} 
where $\eta^{\mu \nu} = \text{diag}(1, -1, -1, -1)$ is the Minkowski metric, $F^{\nu \lambda}$ is the electromagnetic tensor and $\epsilon_{\rho \mu \nu \lambda}$ is the totally antisymmetric tensor such that $\epsilon_{0123}= 1$. 

In order to complete the measurement, we now have to project the position of the particle along the $z$ direction. Formally, this is achieved by defining the two operators $\hat{\Pi}_{+}^{(A)} = \int^{+\infty}_{0} dz_c | z_C \rangle_C \langle z_C |$ and $\hat{\Pi}_{-}^{(A)} = \int_{-\infty}^{0} dz_c | z_C \rangle_C \langle z_C |$, distinguishing whether the particle is respectively  deflected upwards or downwards. For a thorough analysis of a concrete detection of spin via the ``relativistic Stern-Gerlach'' proposed here and more details on the measurement, see Supplemental Information.

The QRF transformation provides the description of the same experiment from the point of view of two different QRFs, which move in a superposition of velocities relative to each other. This treatment of the relativistic Stern-Gerlach experiment makes it possible to associate an operational meaning to the spin of a relativistic quantum particle, thus solving the problem of encoding quantum information in a particle with spin degrees of freedom as in a qubit.

\section{Conclusions}

In this paper, we have provided an operational description of the spin of a special-relativistic quantum particle. Such operational description is hard to obtain with standard methods due to the combined effect of special relativity, which makes the spin and momentum not separable, and quantum mechanics, which makes it impossible to jump to the rest frame with a standard reference frame transformation. We have introduced the `superposition of Lorentz boosts' transformation to the rest frame of a quantum particle, moving in a superposition of relativistic velocities from the point of view of the laboratory. We have found how the state transforms under such quantum reference frame transformation and identified a set of observables in the laboratory frame which satisfies the $\mathfrak{su}(2)$ algebra and has the same eigenvalues as the spin in its rest frame. In addition, this set complies with the desiderata for a relativistic spin operator in Ref. \cite{bauke_relativisticspin}: it commutes with the free Dirac Hamiltonian, it satisfies the $\mathfrak{su}(2)$ algebra, and it has the same eigenvalues as the spin in its rest frame. In addition, it has the correct nonrelativistic limit. It can be easily shown, in fact, that our operator $\vec{\hat{\Xi}}$ coincides with the Foldy-Wouthuysen spin operator \cite{foldywouthuysen, caban_spinobservable}. Thanks to the unitarity of the transformation, probabilities are the same in the rest frame and in the laboratory frame. Finally, we have generalised the Stern-Gerlach to the special-relativistic regime by means of a transformation of the interaction Hamiltonian from the rest frame to the laboratory frame. Such generalisation opens up the possibility of performing quantum information protocols with spin in the special-relativistic regime. 

\acknowledgements{We would like to thank Carlos Pineda for helpful discussions. We acknowledge support from the research platform “Testing Quantum and Gravity Interface with Single Photons” (TURIS), the Austrian Science Fund (FWF) through the project I-2526-N27 and I-2906, the \"{O}AW Innovationsfonds-Projekt ``Quantum Regime of Gravitational Source Masses'', and the doctoral program “Complex Quantum Systems” (CoQuS) under Project W1210-N25. We also acknowledge financial support from the EU Collaborative Project TEQ (Grant Agreement 766900). This work was funded by a grant from the Foundational Questions Institute (FQXi) Fund. This publication was made possible through the support of a grant from the John Templeton Foundation  (Project 60609). The opinions expressed in this publication are those of the authors and do not necessarily reflect the views of the John Templeton Foundation.}

\appendix

\section{Review of the formalism for quantum reference frames in the Galilean case}
\label{App:reviewQRF}

In Ref.~\cite{QRF}, a formalism to describe quantum states, dynamics, and measurements from the point of view of a quantum reference frame (QRF) was introduced. This formalism is operational, in that primitive laboratory operations ---preparation, transformations, and measurements of quantum states--- have fundamental status, and relational, because everything is formulated in terms of relational quantities and the formalism does not require the presence of any external or absolute reference frame. 

The simplest situation is composed of three systems C (the initial QRF), A (the final QRF), and B (a quantum system). Since only the relational degrees of freedom play a role, from the point of view of C the relational degrees of freedom of A and B relative to C are described, and from the point of view of A the relational degrees of freedom of B and C relative to A are described. For instance, if we want to consider relative coordinates, C associates the position operator $\hat{x}_A$ to A, and the position operator $\hat{x}_B$ to B. Operationally, these operators indicate the relative distance between A or B and the origin of the QRF C. From the point of view of A, B has the position operator $\hat{q}_B$ associated to it, while C has the position operator $\hat{q}_C$. The operators in the two QRFs are related via a QRF transformation $\hat{S}_x$, acting as $\hat{S}_x \hat{x}_B \hat{S}_x^\dagger = \hat{q}_B - \hat{q}_C$ and $\hat{S}_x \hat{x}_A \hat{S}_x^\dagger = - \hat{q}_C$. The explicit expression of the transformation to change QRF from C to A is
\begin{equation}
	\hat{S}_x = \mathcal{P}_{AC}e^{\frac{i}{\hbar}\hat{x}_A \hat{p}_B},
\end{equation}
where $\mathcal{P}_{AC}$ is the ``parity-swap'' operator acting as $\mathcal{P}_{AC} \hat{x}_A \mathcal{P}_{AC}^\dagger = - \hat{q}_C$ and $\mathcal{P}_{AC} \hat{p}_A \mathcal{P}_{AC}^\dagger = - \hat{\pi}_C$ and $\hat{\pi}_C$ is the canonically conjugated operator to $\hat{q}_C$. Intuitively, this transformation acts as a ``controlled translation'' on the state of the QRF A. One of the main consequences of this transformation is that the notion of entanglement and superposition is frame-dependent. To illustrate this point, let us assume that C assigns to A and B the quantum state $\left| \Psi \right\rangle_{AB}^{(C)} = \frac{1}{\sqrt{2}} \left( \left| x_1 \right\rangle_A + \left| x_2 \right\rangle_A \right) \left| x_0 \right\rangle_B$, which is separable. After the QRF transformation, the state in A's QRF becomes entangled, i.e. $\left| \Phi \right\rangle_{BC}^{(A)} = \hat{S}_x\left| \Psi \right\rangle_{AB}^{(C)} = \frac{1}{\sqrt{2}} \left( \left| x_0 - x_1 \right\rangle_B\left| -x_1 \right\rangle_C + \left| x_0 - x_2 \right\rangle_B\left| -x_2 \right\rangle_C \right)$.

Another relevant property of the formalism for QRFs introduced in Ref.~\cite{QRF} is that the probability to observe an outcome of a measurement is conserved under change of QRF thanks to the unitarity of the QRF transformation. Specifically, if the probabilities in the QRF of C are calculated as
\begin{equation}
	p (b^*) = \mathrm{Tr}\left[ \hat{\rho}_{AB}^{(C)} \hat{O}_{AB}^{(C)}(b^*) \right],
\end{equation}
where $\hat{\rho}_{AB}^{(C)}$ is the quantum state, $\hat{O}_{AB}^{(C)}$ the operator, and $\hat{O}_{AB}^{(C)}(b^*)$ the projector on a specific outcome in C's perspective, in A's QRF, with analogous notation, we find that
\begin{equation}
	p (b^*) = \mathrm{Tr}\left[ \hat{\rho}_{BC}^{(A)} \hat{O}_{BC}^{(A)}(b^*) \right],
\end{equation}
where $\hat{\rho}_{BC}^{(A)} = \hat{S}_x \hat{\rho}_{AB}^{(C)} \hat{S}_x^\dagger$ and $\hat{O}_{BC}^{(A)} = \hat{S}_x \hat{O}_{AB}^{(C)} \hat{S}_x^\dagger$.

Notice that in this work we use the terminology ``superposition of Lorentz boosts'' because we describe physics from the point of view of two QRFs moving in a superposition of relativistic velocities relative to each other. This is different to the definition of such transformations given in Ref.~\cite{QRF}, where the transformation is written as the classical reference frame transformation with the parameter of the transformation promoted to an operator, because the transformation of the spin degree of freedom alone does not correspond to a Lorentz boost. Notice, however, that the QRF transformation would be completely analogous to those in Ref.~\cite{QRF} if we considered a Klein-Gordon field instead of the spin $\tilde{A}$. In this case, the QRF transformation would be given by a Lorentz boost, with the parameter $v$ replaced by a function of the operator $\hat{\pi}_C$, followed by a generalised parity-swap operator.

\section{Derivation of the quantum reference frame transformation}
\label{App:QRFTrans}

The QRF transformation $\hat{S}_L$ can be derived by considering a tripartite Hilbert space of the external degrees of freedom of the particle (A), the spin of the particle ($\tilde{A}$), and the laboratory (C). A general basis element, in the rest frame of the particle A is $|k_A; \vec{\sigma} \rangle_{A\tilde{A}} | \pi\rangle_C$, where $k_A^\mu = (m_A c,\vec{0} )$. If we wish to change to the QRF of the laboratory C, we need to apply a standard Lorentz boost to the state of the particle A$\tilde{A}$, where the boost parameter is controlled by the momentum of the laboratory $\hat{\pi}_C$, and then boost the laboratory state by the momentum of A. The full transformation reads
\begin{equation}
	\hat{S}_{ext} = \hat{U}_C^\dagger (L_{\frac{m_C}{m_A}\hat{p}_A})\hat{U}_{A\tilde{A}}(L_{-\frac{m_A}{m_C}\hat{\pi}_C}),
\end{equation}
where $\hat{U}_X(L_{\hat{p}_Y})$ (with $X\neq Y$) is the standard Lorentz boost by the momentum operator $\hat{p}_Y$ (i.e., the standard Lorentz boost where the parameter $p$ has been promoted to the operator $\hat{p}_Y$), acting on the Hilbert spaces $X$ and $Y$ as $\hat{U}_{A\tilde{A}}(L_{-\frac{m_A}{m_C}\hat{\pi}_C}) |k_A; \vec{\sigma} \rangle_{A\tilde{A}} | \pi\rangle_C = |-\frac{m_A}{m_C}\pi; \Sigma_\pi \rangle_{A\tilde{A}} | \pi\rangle_C$ and $\hat{U}_C^\dagger (L_{\frac{m_C}{m_A}\hat{p}_A}) |-\frac{m_A}{m_C}\pi; \Sigma_\pi \rangle_{A\tilde{A}} | \pi\rangle_C = |- \frac{m_A}{m_C}\pi; \Sigma_\pi \rangle_{A\tilde{A}} | k_C \rangle_C$. 
The action on a basis of the total Hilbert space is
\begin{equation}
	\hat{S}_{ext} |k_A; \vec{\sigma} \rangle_{A\tilde{A}} | \pi\rangle_C = | -\frac{m_A}{m_C}\pi; \Sigma_\pi \rangle_{A\tilde{A}} | k_C\rangle_C,
\end{equation}
with $k_C^\mu = (m_C c,\vec{0})$. We notice that the Hilbert spaces of the QRF, A and C respectively before and after the transformation, are irrelevant and can be omitted, as they are only labelled by the zero momentum $4$-vector $k_A$ and $k_C$ and represent no dynamical degrees of freedom (their quantum states do not change in time). Therefore, in accordance with the QRF formalism introduced in Ref.~\cite{QRF}, we can drop them and write the QRF transformation in the more compact form $\hat{S}_L$ in the main text
\begin{equation}
	\hat{S}_L | \vec{\sigma} \rangle_{\tilde{A}} | \pi\rangle_C = | -\frac{m_A}{m_C}\pi; \Sigma_\pi \rangle_{A\tilde{A}}.
\end{equation}
This transformation can be explicitly written as
\begin{equation}
	\hat{S}_L = \mathcal{P}_{AC}^{(v)} \hat{U}_{\tilde{A}}(\hat{\pi}_C),
\end{equation}
where $\hat{U}_{\tilde{A}}(\hat{\pi}_C)$ is a unitary operator acting on the joint Hilbert space $\mathcal{H}_{\tilde{A}}\otimes \mathcal{H}_C$, and $\mathcal{P}_{CA}^{(v)}= P_{AC} \exp \left(\frac{i}{\hbar}\log \sqrt{\frac{m_A}{m_C}}(\hat{q}_C \hat{\pi}_C + \hat{\pi}_C \hat{q}_C)\right)$ is the `generalised parity-swap operator' introduced in Ref.~\cite{QRF}. Here, $P_{AC}$ is the parity-swap operator mapping $\hat{x}_A \rightarrow -\hat{q}_C$ and $\hat{p}_A \rightarrow -\hat{\pi}_C$ (and viceversa). Additionally to the action of $P_{AC}$, the operator $\mathcal{P}_{AC}^{(v)}$ rescales the momentum of A by the ratio of the masses of A and C, i.e., $\mathcal{P}_{AC}^{(v)} \hat{p}_A \mathcal{P}_{AC}^{(v)\dagger} = - \frac{m_A}{m_C} \hat{\pi}_C$, such that the velocity operators of A and C are mapped as $\hat{v}_A \mapsto - \hat{v}_C $.

\section{Action of the spin operators}
\label{App:LorentzSpin}

The spin of a particle has a natural definition in the rest frame. In a different reference frame, quantum field theory predicts that a more general quantity, the total angular momentum, is conserved, and spin alone is no longer operationally well defined, because the splitting of the angular momentum and spin momentum is not unique. However, it is possible to associate a $4$-vector to the spin operator, the Pauli-Luba\'{n}ski spin operator, which can then be transformed in a covariant way. The treatment can be found, e.g., in Refs.~\cite{jackson_electrodynamics, sexlurbantke}. In the rest frame, the components of the covariant spin are $\hat{\sigma}^\nu = (\hat{0}, \hat{\sigma}_x, \hat{\sigma}_y, \hat{\sigma}_z)$, where $\hat{\sigma}_i$, $i=x,y,z$, are the Pauli operators, which generate the group $SU(2)$. We now want to boost the state of the particle to a reference frame that moves in general in a superposition of velocities. To this end we introduce the momentum operator $\vec{\hat{p}} = m \gamma \vec{\hat{v}}$, where $m$ is the mass of the particle, $\hat{\gamma} = \sqrt{1+ \frac{\vec{\hat{p}}^2}{m^2c^2}} = \left(1-\frac{\vec{\hat{v}}^2}{c^2}\right)^{-1/2}$ and $\vec{\hat{v}}$ is the velocity operator of the particle in the new reference frame. We define by $U(L_{-\hat{p}})$ the operator representing a pure Lorentz boost to the new reference frame, where the boost operator $L_{\hat{p}}$ is explicitly written as the matrix
\begin{equation} \label{eq:Lp}
	L_{\hat{p}} =  \left(\begin{array}{c|c}
			{} & {}\\
			\frac{{\hat{p}}_0}{mc} & -\frac{{\hat{p}}_i}{m c}\\
			{} & {}\\
			\hline 
			{} & {}\\
			-\frac{{\hat{p}}_i}{m c} & \delta_{ij} + \frac{{\hat{p}}_i {\hat{p}}_j}{m c ({\hat{p}}_0 + m c)}\\
			{} & {}
		\end{array}\right),
\end{equation}
where $i,j=x,y,z$. The Pauli-Luba\'{n}ski spin operator is defined (up to a factor $m_A$) as $\hat{\Sigma}_{\hat{p}}^\mu= \hat{U}^\dagger (L_{\hat{p}}) \hat{\sigma}^\mu \hat{U}^\dagger (L_{\hat{p}})= (L_{-{\hat{p}}})^\mu_\nu \hat{\sigma}^\nu$ \cite{sexlurbantke}. Formally, this relation can be derived via the action on a generic basis element $\hat{U}(L_p) | k, \vec{\sigma}\rangle = | p, \Sigma_p\rangle$, where $| k, \vec{\sigma}\rangle = \sum_{\lambda}c_\lambda \,| k, \lambda\rangle$ is represented in some spin basis, and $\hat{U}(L_p)$ is the standard Lorentz boost from the rest frame to the frame where the particle has momentum $p$. We can now write the action of the Pauli-Luba\'{n}ski operator as
\begin{equation}
	\begin{aligned}
		\hat{\Sigma}_{\hat{p}}^\mu \, | p, \Sigma_p\rangle &= \hat{\Sigma}_p^\mu \hat{U}(L_p) \, | k, \vec{\sigma}\rangle = \hat{U}(L_p) \hat{U}^\dagger (L_p)\hat{\Sigma}_p^\mu \hat{U}(L_p) \, | k, \vec{\sigma}\rangle=\\
		&= \hat{U}(L_p) (L_{-p})^\mu_\nu \hat{\Sigma}_p^\nu | k, \vec{\sigma}\rangle = \sum_{\lambda}c_\lambda \hat{U}(L_p) (L_{-p})^\mu_\nu \hat{\sigma}^\nu  \,| k, \lambda\rangle = \\
		&= \sum_{\lambda, \lambda'}c_\lambda \hat{U}(L_p) (L_{-p})^\mu_\nu \left[\sigma^\nu\right]_{\lambda' \lambda}  \,| k, \lambda'\rangle = \sum_{\lambda, \lambda'}c_\lambda (L_{-p})^\mu_\nu \left[\sigma^\nu\right]_{\lambda' \lambda}  \,| p, \Sigma_p(\lambda')\rangle=\\
		&= (L_{-p})^\mu_\nu \hat{\sigma}^\nu  \,| p, \Sigma_p\rangle.
	\end{aligned}
\end{equation}
Thus, by making use of Eq.~\eqref{eq:Lp}, it is immediate to verify that
\begin{equation} \label{eq:CovSpin}
	\hat{\Sigma}_{\hat{p}}^0 = \hat{\gamma} \vec{\hat{\beta}} \cdot \vec{\hat{\sigma}}; \qquad \vec{\hat{\Sigma}}_{\hat{p}}= \vec{\hat{\sigma}} + \frac{\hat{\gamma}^2}{\hat{\gamma} + 1}(\vec{\hat{\beta}} \cdot \vec{\hat{\sigma}})\vec{\hat{\beta}},
\end{equation}
where $\vec{\hat{\beta}}= \frac{\vec{\hat{v}}}{c} = \frac{\vec{\hat{p}}}{\sqrt{m^2c^2+|\hat{p}|^2}}$. 

Notice that the introduction of the fourth spin component does not add degrees of freedom, because the $4$-vector has to satisfy the covariant constraint $\eta_{\mu \nu}\hat{p}^\mu \hat{\Sigma}^\nu_{\hat{p}}=0$, where $\eta_{\mu \nu} = \text{diag}(1, -1, -1, -1)$ is the Minkowski metric. With an analogous method it is possible to show that $\hat{\Xi}^i \, | p, \Sigma_p\rangle = \hat{U}(L_p) (\mathbb{1}\otimes \vec{\sigma})\hat{U}^\dagger (L_p) \, | p, \Sigma_p\rangle = (L_p)^i_\mu \hat{\Sigma}_p^\mu \, | p, \Sigma_p\rangle$. Thus, the action of the two operators $\hat{\Sigma}_p$ and $\hat{\Xi}$ is obtained by Lorentz-transforming the other. 
 
 Formally, the total Hilbert space can be described as a fiber bundle, the base manifold being the space of square-integrable functions and the fibers being the two-dimensional Hilbert space $\mathcal{H}_p$, representing the spin degree of freedom. 

\section{Concrete analysis of the relativistic Stern-Gerlach}
\label{App:concrete}

Let us consider an experiment performed in the laboratory frame C. We describe the $x$ and $z$ external degrees of freedom of particle A and the spin degrees of freedom $\tilde{A}$, and take the state at time $t=0$ to be
\begin{equation} \label{eq:instateEx}
	\left| \Psi_0 \right>_{A\tilde{A}}^{(C)} = \cos \theta \left| \Psi_0^+ \right>_{A\tilde{A}} + \sin \theta \left| \Psi_0^- \right>_{A\tilde{A}},
\end{equation}
where 
\begin{equation}
	\left| \Psi_0^{\pm} \right>_{A\tilde{A}} = \left| \psi_0^z \right>_{A} \left| \phi_x^{\pm} \right>_{A\tilde{A}}
\end{equation}
and $\left| \psi_0^z \right>_{A} = \int dp_z \psi_z (p_z) \left| p_z \right>_{A}$ with $\psi_z (p_z) = \frac{1}{(2\pi s_z^2)^{1/4}} e^{-  \frac{p_z^2}{4 s_z^2}}$ being a gaussian wavepacket centered in zero with standard deviation $s_z$, and $\left| \phi_x^{\pm} \right>_{A\tilde{A}} = \int d \mu( p_x) \phi_x(p_x) \left| p_x ; \Sigma_{p_x}^{\pm} \right>_{A\tilde{A}}$. Here, $\phi_x(p_x)$ is a general wavepacket and $\left| p_x ; \Sigma_{p_x}^{\pm} \right>_{A\tilde{A}}$ is the eigenvector with eigenvalue $\lambda_{\pm} = \pm 1$ of the operator $\hat{\Xi}_z$. As in the main text, we assume that the motion along $z$ is nonrelativistic. In the laboratory frame, we engineer an interaction between the system A and the magnetic field such that the interaction Hamiltonian is $\hat{H}_{int}^{(C)}= \mu B^{(C)}_z(z) \hat{\Xi}_z$, with $B^{(C)}_z(z)= B^0_z - \alpha z$, $\alpha>0$. Evolving the state in the interaction picture, we get
\begin{equation}
	\left| \Psi_t \right>_{A\tilde{A}}^{(C)} = e^{-\frac{i}{\hbar} \hat{H}_{int}^{(C)} t} \left| \Psi_0 \right>_{A\tilde{A}}^{(C)}= \cos \theta \left| \Psi_t^+ \right>_{A\tilde{A}} + \sin \theta \left| \Psi_t^- \right>_{A\tilde{A}},
\end{equation}
where $\left| \Psi_t^{\pm} \right>_{A\tilde{A}}= \int d\mu(p_x) dp_z \phi_x (p_x) \psi_z (p_z) e^{\mp\frac{i}{\hbar}B^0_z t}e^{\pm\frac{i}{\hbar}\alpha \mu t \hat{z}_A} \left| p_z \right>_{A} \left| p_x ; \Sigma_{p_x}^{\pm} \right>_{A\tilde{A}}$. This state can be rewritten as $\left| \Psi_t^{\pm} \right>_{A\tilde{A}}= e^{\mp\frac{i}{\hbar}B^0_z t}\int d\mu(p_x) dp_z \phi_x (p_x) \psi_z (p_z \mp p^*(t)) \left| p_z \right>_{A} \left| p_x ; \Sigma_{p_x}^{\pm} \right>_{A\tilde{A}}$, where $p^*(t) = \frac{\alpha \mu t}{\hbar}$. Hence, under the effect of the interaction with the magnetic field, the gaussian wavepacket along $z$ gets split into two wavepackets, moving in opposite directions according to the state of the spin. The two gaussians become distinguishable under the condition $\left| {}_{A}\left\langle \psi_z^+ | \psi_z^- \right>_{A} \right| \ll 1$, where we have defined $\left| \psi_z^{\pm} \right>_{A} = \int dp_z \psi_z (p_z \mp p^*(t))\left| p_z \right\rangle_{A}$. This condition is satisfied if $t > \frac{\hbar s_z}{\alpha \mu}$ (neglecting factors of order one). If we now define the projectors $\hat{\Pi}^{\pm}_z = \left| \psi_z^{\pm} \right>_{A} \left\langle \psi_z^{\pm} \right|$, we can calculate the probabilities to find spin ``up'' or spin ``down'' as
\begin{equation}
	p_\pm = \text{Tr}\left[ \left| \Psi_t \right\rangle_{A\tilde{A}}^{(C)} \left\langle \Psi_t \right| \hat{\Pi}^{\pm}_z \right],
\end{equation}
where $\left| \Psi_t \right\rangle_{A\tilde{A}}^{(C)}$ denotes the time-evolved state of Eq.~\eqref{eq:instateEx}. Under the condition $t > \frac{\hbar s_z}{\alpha \mu}$ we find that the only relevant contributions to the probabilities lead to $p_+ = \cos^2 \theta$ and $p_- = \sin^2 \theta$. 

Note that the time at which the projectors $\hat{\Pi}^{\pm}_z$ are applied also transforms with a ``superposition of Lorentz boosts'' between different QRFs. This means that, if we choose a time to apply the measurement in a specific reference frame, the corresponding measurement will take place in a superposition of times in a different QRF. This happens because of the relation between the proper times $t_C = \hat{\gamma}_A t_A$ and the fact that the particle moves in a superposition of velocities. Therefore, measuring the time of arrival of the particle would provide information about the momentum of the particle.

\end{document}